# Title: Tunguska cosmic body of 1908: is it from planet Mars?

**Authors:**

John Anfinogenov,[1] Larisa Budaeva,[2] Dmitry Kuznetsov,[3] and Yana Anfinogenova[2,4]*

**Affiliations:**

[1]Tungussky Nature Reserve, Ministry of Natural Resources and Ecology of the Russian Federation.

[2]National Research Tomsk State University, Ministry of Education and Science of the Russian Federation

[3]National Research Tomsk Polytechnic University, Ministry of Education and Science of the Russian Federation

[4]Tomsk University of Control Systems and Radioelectronics, Ministry of Education and Science of the Russian Federation

*Corresponding author: Dr. Yana Anfinogenova. Address: TUSUR, FIT, 40 Lenina Prospect, Tomsk, 634050, Russia. Tel: +79095390220. E-mail: anfiyj@gmail.com

Abstract word count: 249

Number of References: 27

Number of Tables: 1

Number of Figures: 5

**Abstract:** The aim of the study was to discover remnants of the 1908 Tunguska meteorite. Main objective of the field studies was identification of exotic rocks, furrows, and penetration funnels reported by the first eyewitnesses, residents of the area with severe forest destruction. Main methods included decoding of aerial survey photographs, systematic survey of the epicenter area of the Tunguska explosion, exploratory excavations of the objects of interest, reconstruction studies of exotic boulder by using its splinters, mineralogical and spectral analysis of specimens, experimental attempt of plasma-induced reproduction of the fusion crust on specimen. The authors present results on discovery of penetration funnel-like structures; exotic boulder (known as John's Stone)) with its shear-fractured splinters and fresh furrow in the permafrost; several splinters with glassy coatings; evidence of high-speed John's Stone deceleration in the permafrost; and clear consistency in geometry of spacial arrangements of all splinters, furrows, cleaved pebbles according to data of reconstruction studies. John's Stone is composed of highly silicified gravelite sandstone (98.5% $SiO_2$) with grain size of 0.5 to 1.5 *cm*. Outer surface of several splinters showed continuous glassy coating similar to shiny fusion crust reminiscent of freshly applied enamel. Plasma-induced heating of John's Stone specimen led to its explosive disintegration; residue presented with whitish semi-transparent pumice-like grains and irregularly shaped fused particles. Overall, our data suggest that John's Stone is Tunguska meteorite candidate. Recent discoveries of sedimentary rocks, lithified gravelite sandstone, clay, and quartz on Mars provide rationale for search and identification of silica-rich Martian meteorites of sedimentary origin.

## Introduction:

In history of Tunguska catastrophe (June 30, 1908) studies, there is an enigmatic chapter about a search for stones, which, according to eyewitnesses' reports, appeared after the catastrophe (Vasiliev et al. 1981). Eyewitnesses reported about the stone, which looked like a couchant animal ("Deer Stone") and appeared "from nowhere" amidst the catastrophe forest far from any rock outcroppings after the Tunguska catastrophe. Tungus (Evenkis) whose name was Daunov saw big shiny stone of tin color. Vitaliy Voronov reported about stone of gray color with sharp edges, which his grandmother kept in her coffer: the woman stated that the stone was removed from a tree trunk after the catastrophe of 1908. Nastya Dzhonkoul testified about seeing a rock falling from the sky: "A big stone, as big as a deerskin tent, jumped about the ground twice or trice and then swamped in the bog" (Suslov 1967). Several local Evenkis reported about fresh furrows in the epicenter with stones in the furrow walls. According to their reports, those stones looked like "quartz crystals". For long time, attempts to find those stones were unsuccessful. However, it would be unreasonable to disregard Evenkis' reports because several large geological structures such as craters and furrows mentioned by Evenkis have been indeed found later. In 1930, Constantine Yankovsky found and photographed a strange stone ("Yankovsky's Stone") in the epicentral area. The stone looked like a stony meteorite with regmaglypts. However, the very that day, Yankovsky was bitten by viper and had to take care of himself to survive; the stone was lost. The dimensions of Yankovsky's stone were over 2 x 1 x 1 *m*.

Many speculations have been suggested for the origin of Tunguska meteorite though it appears that the Tunguska event represents a typical fate for stony asteroids tens of meters in radius entering the Earth's atmosphere at common hypersonic velocities (Chyba et al. 1993). In *Nature News* article, Mark Peplow (2013) referenced a paper published in May, 2013 by Andrey Zlobin (2013) who claimed to have found pebbles from the Tunguska meteor. The article describes three specimens, one of which consists of quartz-like substance and has visible traces of melting on the surface. According to Mark Peplow, Zlobin's paper "was quickly dismissed by field specialists". Indeed, quartz has been considered a minor to rare mineral in any type of meteorite identified so far (Grady 2000). However, we believe that the traces of melting on a surface of the rock from the epicentral area of the Tunguska catastrophe may provide rationale for meticulous examination of the specimen. The quartz-rich specimen seems of interest to us due to recent discoveries of quartz on Mars and in connection with our own field discoveries in Tunguska. In this paper, for the first time in literature written in English language, we present data on our candidate for Tunguska meteorite and discuss it in the context of the latest discoveries of quartz and sedimentary rocks on Mars.

## Methods:

**Analysis of the first eyewitness' reports and decoding of aerial survey photographs.** Analysis of the first reports of the Tunguska catastrophe eyewitnesses was performed in 1960s and focused on identifying fresh geomorphological formations (holes, furrows, and exotic rocks). Aerial survey photographs of 1938 and 1949 were decoded from 1964 to 1972. Decoding consisted in viewing the stereoscopic pairs of photographs with aim to identify fresh geomorphological formations. Aerial survey photographs of 1949 were used to determine configuration of the complete forest fall area; it was necessary to establish scattering ellipse far end with more massive meteorites.

**Field Studies.** Field studies took place in 1965-1968, 1972-1974, 1978-1982, 1988-1989, and 1995-2002. Research *in situ* was done within the areas of interest identified based on the analysis of the first eyewitness' reports and decoding of aerial survey photographs. Groups of two to five volunteers were combing the area by walking along the designated parallel routes.

**Exploratory excavations.** In 1965 to 1968, penetration funnel-like structures ($n$ = 30) were found in the area of interest. Exploratory excavations of funnels ($n$ = 3) consisted in cutting the narrow longitudinal tranches along the diameter of funnel in direction parallel to azimuth of the Tunguska body flight. High-silica splinters, recovered from the funnels, were presented to certified geologists for identification. The rest of the funnels were left undisturbed.

On July 19, 1972, exotic boulder (John's Stone (JS)) was discovered at the Stoykovich Mountain. Moss cover was carefully removed from the boulder. Five short radial tranches were excavated with aim to find splinters of the boulder. Then, the 0.5 m-deep circular trench of 10 m in radius was excavated around the boulder. Upon identification of the narrow sector with high content of splinters, transverse 0.5-*m*-wide 2.5-*m*-deep trenches were excavated at distances of 5 *m* and 10 *m* from the center of the boulder. Excavations revealed buried impact furrow whose configuration, depth, width, azimuth, and composition of the filling ground were documented. Three-dimensional localizations of all recovered splinters were documented. Beginning part of the furrow was preserved undisturbed for further studies.

**Reconstruction studies.** Upon finding numerous splinters of JS, reconstruction studies were done. In case of large splinters, reconstruction was achieved by identifying their initial place on John's Stone according to configuration of their surfaces and characteristic lithological patters including positions of half-split pebbles which belonged to both the main boulder and splinters. In case of smaller plate-like shear splinters, we reconstructed continuous flat puzzle-like sheets.

**Mineralogical and spectral analysis.** Lithological, mineralogical, and spectral analysis of specimens was performed in the State Enterprise “Krasnoyarsk Geological Survey”. Lithological and mineralogical characteristics were studied according to standard methodology. Spectral analysis was done by using high-performance automated spectrometer ДФС-8 (Russia) with photoelectronic multipliers (Hamamatsu, Japan). Results are presented as percent concentration of chemical elements.

**Experimental reproduction of fusion crust on John's Stone specimen.** To reproduce fusion crust on a specimen of JS, the sample was heated with plasma beam (5000 °C) in a chamber of electric arc plasma generator in the Department of Silicates at National Research Tomsk Polytechnic University).

**Estimation of energy for destruction of permafrost ground.** According to direct dimensional measurements, total volume of permafrost destroyed and displaced by JS was about 50 $m^3$. Compressive strength of permafrost was equated to the strength of moderately strong concretes (limit of strength of about 50 $MN/m^2$). This supposition was accepted based on comparison of the frozen cement grouts with cement grouts hardened at normal technological conditions. The landing velocity of the meteoroid was calculated via the equation for its kinetic energy which was equal to work on destruction of the given volume of permafrost.

**Previous dissemination of JS specimens:**

Oral reports and some characteristic specimens of JS were shared with experts from the State Enterprise “Krasnoyarsk Geological Survey” (Krasnoyarsk, Russia), Tomsk State University (Tomsk, Russia), National Research Tomsk Polytechnic University (Tomsk, Russia), Committee on Meteorites of the USSR Academy of Sciences (Moscow, Russia), Natural History Museum (London, UK), University of Bologna (Bologna, Italy), State University of New York (Albany, NY, USA), and some others.

**Results:**

During the expeditions of 1965 to 1968, John Anfinogenov's group discovered funnel-like structures (*n* = 30) of 1.0 to 2.5 *m* in diameter with typical raised rims in the epicenter area of the Tunguska meteoroid explosion (Figure 1). Exploratory excavations were done in three funnels; two of them contained shear-fractured splinters (*n* = 2) composed of strongly silicified sandstone. Geologists stated that these specimens “may not be meteorites” and called the penetration funnels-like structures “old anthills”. The specimens were eventually lost; the penetration funnels were disregarded.

**John's Stone. Reconstruction studies.** In 1972, the group discovered unusual rock (John's Stone (JS)) on the top of the Stoykovich Mountain (Figure 1). The first impression was that the rock was the so-called “Deer Stone” or “ Yankovsky's Stone”, reported by Evenkis and Constantine Yankovsky, respectively. The boulder located 30 meters from the bottomland swamp near the east foot of the west uplift of the Stoykovich Mountain, five meters from the east taxator profile. John's Stone was hardly visible at first. Only a small patch (20 x 20 *cm*) of unusual white-gray-blue color with evident scorch marks peeked out of slightly raised moss-covered hillock. The patch was oriented south-east and raised 35 *cm* above the surrounding ground. Young 14-year-old 1.5-*m*-high birch grew on the top of the hillock. Nearly all surface of JS was covered by 2—3-*cm*-thick layer of moss, fallen leaves, and interlacing roots. Birch roots were clinging to the boulder following the cavities on its surface down to the ground. Mineral residue in the cavities was insignificant: from 1 to 5 *mm*. Young birches, larch-trees, pines, and 45-year-old junipers grew near JS. When the young birch growing at the top of the hillock was bent by hand, the boulder, covered only with moss, easily came out. All surface of uncoated JS had the same white-gray-blue color. On closer examination, the stone was composed of mostly light-toned concrete-bound pebbles and sand (conglomerate gravelite sandstone) with unwashed scorch marks. The grain size was in a range of 0.5 to 1.5 *cm*, rarely up to 5 *cm*.

Sizes of the above-ground part (cupola) of JS were 2 x 1.5 x 0.5 *m* (Figure 3A); the boulder set along a meridian (from south to north). Cupola was slightly sloping to the north and west and had steeper incline to the east and south. The vertical south wall of the boulder seemed broken. Overall, JS was rounded, almond-shaped. Sizes of JS after complete recovery: 2.5 x 1.7 x 1.2 *m;* estimated mass exceeded 10,000 *kg*. An attempt to split off a sample showed that the boulder was uniformly resistant, concrete, extremely hard, and sonorous. Shears of JS revealed clear quartz, crystalline quartz, and analcime crystals filling space between the pebbles. The north-east, east, south, and upper surfaces of JS cupola had pebbles that were cut and sheared by powerful mechanical shear action; shears were very clean and fresh. The west surface of JS had areas that looked like very hard slags of the same composition (silica), but they were significantly more porous and did not have shears.

On closer examination of the ground surrounding the north part of the boulder, shear-fractured splinter (25 x 18 *cm*) was found. Reconstruction study confirmed that the splinter fitted JS; the splinter was displaced from the corresponding depression on the boulder by 29 *cm* upward. Shears were fresh.

During the season of 1972, north, south, west, and east exploratory trenches were excavated around JS (Figure 2). The north trench contained a large fragment (0.5 *m* in size; 40 to 50 *kg*) near JS. Space between JS and this splinter was filled with hardpack soil. Shears were fresh. The thickness of the specimen ranged from 0.5 to 15 *cm*. Texture of the exposed (north) surface of the splinter differed from that of the other surface: it was more smoothed and weathered. The color of the shear-fractured surfaces differed from the rest of surfaces on both JS and the splinter. Reconstruction study confirmed that this splinter perfectly fitted the north flat side of JS.

Small chipped plate-like splinters were found in small radial exploratory trenches at a distance of 1 to 1.20 *m* from JS. The shear splinters were found **only** in the trenches excavated eastward from the north-east part of the boulder.

Large circular exploratory trench (10-*m* in radius; 40-*cm*-deep; 40-*cm*-wide) was excavated in the ground ringing about JS (Figure 2). Sear splinters were found **only** in the east sector of the circular trench along the azimuth of 85°—100° (Figure 2).

Transverse exploratory trench (1.5 *m* deep) was excavated at a distance of 6 *m* from JS. Cross sections on the trench walls revealed V-shaped furrow (described in more detail below) filled with loose black ground (Figure 3B). The walls of the furrow contained numerous splinters of JS; the splinters were imbedded into the furrow walls; they were also found in turf cover on ground surface.

A 0.5-*m*-diameter spot of light-toned sand (Figure 3C&D) was found under the litterfall in the east sector of the circular exploratory trench at a distance of 10 *m* from JS along the azimuth of 85°. The accumulation of sand contained extraordinary plate-like splinters that fitted each other. Two of these splinters had **continuous well-formed glassy coating that looked like shiny fusion crust reminiscent of freshly applied enamel**. Only one side of these plate-like specimens had glassy coating; other sides were freshly shear-fractured. Above the spot of the light-toned sand, there was uprooted larch-tree without soil between the roots; the tree crown was directed north-east.

At depth of > 1.5 *m* and 0.5 *m* aside of the above mentioned spot of sand, massive (2,000 kg) splinter (Satellite Stone (SS)) was found (Figure 2; Figure 3E). Mutual positional relationship of pebbles suggested that SS broke off the south wall of JS (Figure 4). Satellite Stone was overturned upside down relative to JS. Sand layer stretched up to the ground surface north-west from SS: upper part of sand layer contained the above mentioned specimens with glassy coating.

Splinters, found in the north wall of the furrow, accurately fitted corresponding places on the upper part of JS. At a distance of 9 *m* from JS, the south wall of the furrow contained massive splinter accurately matching the bottom side of JS.

Importantly, roundish plate-like splinter (30 *cm* in diameter and 10 *cm* thick) was recovered from under the south side of SS. We named this specimen “Red Riding Hood” (RRH) due to reddish color of its coating. We believe that reddish color of RRH was caused by its years-long-contact with the ground saturated with iron oxides present at the bottom part of the furrow under SS (possible replacement by terrestrial minerals). Reconstruction study showed that RRH accurately fitted JS cupola.

Number of specimens, found in the exploratory trenches, exceeded 350. Mostly, these were little scaly plate-like shear-fractured splinters. Sizes ranged from several centimeters in diameter and several millimeters in thickness to thirty centimeters in diameter and several centimeters in thickness. Some of them had ideally smooth glassy outer surfaces. We were able to reconstruct two continuous flat puzzle-like sheets made of dozens of plate-like shear-fractured splinters. Total area of these sheets was about one fourth of a square meter; splinters perfectly gapless fitted each other. Surface color varied from light--bluish-gray to dark rusty-brown (Figure 5). We left preserved areas around JS non-excavated for further studies.

**Furrows.** Exploratory excavations revealed furrow south-east of SS. The furrow was slightly sloping deeper into the ground. Further excavations exposed the second hidden buried furrow (1.85—1.90 *m* deep) which extended the furrow from SS to the very JS (Figure 2). This furrow had very definite fractured boundaries/edges between the light-toned frozen sand and loose black organic-matter-containing soil filling the furrow. This black soil was very loose and easy to excavate: metallic pole poked down through 1.2 *m* of this soil without effort. Furrow depth slightly declined toward JS. Cross section configuration of the furrow perfectly matched the profile of JS on the edge (Figure 3B&E). Transverse exploratory tranches, excavated across the furrow, revealed numerous variable-size splinters whose surfaces accurately matched certain places on JS. The walls of the large furrow contained quartz pebbles (up to 10 *cm* in diameter) which were half-cleaved; the halves matched each other.

Based on the direction of the furrows, trajectory of JS landing was estimated. Magnetic azimuth of the entry into the ground was 280°—320°; projection of the trajectory was 100°—140°. Pattern of permafrost destruction indicated high-speed entry and lateral ricochet of JS in the ground with further deceleration and breakage. Since mass of JS is known (about 10,000 kg), we estimated the landing velocity of the meteoroid via the equation for its kinetic energy which was equal to work on destruction of the given volume of permafrost ($\approx 50\ m^3$).

$$E_{Entry} \cong \sigma_{Strength} \times V_{Ground}$$

According to our calculations, the JS landing velocity was $\approx$ 550 *m*/c.

**Lithological and spectral analysis.** John's Stone is composed of sedimentary metamorphic quartz-rich conglomerate gravelite sandstone with grain size of 0.5 to 1.5 *cm,* rarely up to 5 *cm*. The rock consists of 98.5% of $SiO_2$ (80—85% of silica materials and minerals; 15—20% of siliceous cement). Mineral composition includes quartz and sparse analcime; some experts noted lechatelierite. Mineralogical composition, except lechatelierite, is hardly distinguishable from terrestrial abyssal sedimentary rocks. John's Stone is extremely hard, sonorous, uniformly resistant, and high-strength rock. Petrographic analysis showed that glassy coatings on some splinters are fine-crystalline structures of the same composition as the adjoining grains. Thickness of this crust is up to 0.5 *mm*. Table I shows the results of spectral analysis of John's Stone specimen.

We were unable to reproduce fusion crust on JS specimen. Surface heating by plasma beam (5000 °C) in a chamber of electric arc plasma generator led to the explosive disintegration of the specimen. Discolored residue was composed of whitish semi-transparent pumice-like grains and irregularly-shaped fused particles; grain size ranged from 1 to 5 *mm*.

**Discussion:**

John's Stone locates on quaternary deposits at the top of the Stoykovich Mountain. Large light-toned sedimentary boulders made of strongly silicified sandstone are exotic to the bolson of the Stoykovich mountain whose picks and shoulders are composed of traps (anogenic, igneous rocks, Tunguska basalts). There are no signs of past glaciation throughout the region in-between the Podkamennaya Tunguska and Nizhnyaya Tunguska rivers (Parmuzin 1956). Decoding of aerial survey photographs covering area within 40 km from the epicenter demonstrates that geological formations like active diatremes are unlikely in the region. We were unable to reproduce fusion crust on JS specimen: surface heating by plasma beam led to the explosive disintegration of the specimen. This result may suggest that the crystal structure of minerals in the specimen was overstressed due to previous history of high-energy events. The idea that JS has cosmic origin is consistent with the discovery of significantly higher content of glassy silica microspherules in peat layer of 1908 (Dolgov et al. 1973). Peat layer of 1908 contained up to hundredfold-higher count of gray and colorless transparent silica microspherules than the adjacent peat layers. Data of neutron activation analysis showed that chemical composition of microspherules was distinct from that of terrestrial rocks, industrial glass, known stony meteorites, tektites, and Moon rocks (Kolesnikov et al. 1976). We suggest that JS has high similarity, but not relatedness, to the terrestrial rocks.

Martian meteorite existence has been recognized since early 1980s (Bogard and Johnson 1983).[10] They display igneous mineral composition, characteristic isotope ratios, and specific gas and water inclusions. Of over 61,000 meteorites that have been cataloged on Earth, 122 were identified as Martian (as of June 29, 2013) (The Meteoritical Society). Our candidate for Tunguska meteorite is composed of 98.5% $SiO_2$ and has sedimentary origin. Available literature suggest that quartz-bearing deposits are present and consistently co-located with hydrated silica on Mars (Smith and Bandfield 2012). The quartz formed as a diagenetic product of amorphous silica, rather than as a primary igneous mineral. Crystalline quartz is present in isolated exposures in impact craters (Bandfield et al. 2004) and in nearby knobby and fractured terrains (Bandfield 2006) in Antoniadi Crater region. All detected quartz is co-located with plagioclase feldspar implying that these exposures may represent a felsic pluton that was later excavated by impacts (Smith and Bandfield 2012; Bandfield 2004). Antoniadi Crater once contained rivers and lakes (Bridges 2008). Jerolmack D. J. (2013) provided evidence for pebbles on Mars. Williams R. M. E. *et al*. (2013) reported about discovery of conglomerates on Mars — pebbles mixed with sand and turned to rock — resulting from ancient river deposits. Other authors also stated presence of indurated and lithified dunes of sand, sandstone, and clay on Mars (Edgett and Malin 2000; Kerber and Head 2012; Michalski 2013). Similarly to JS from the epicenter of the Tunguska catastrophe, aeolian ridges on Mars are light-toned (Kerber and Head 2012). Outcrops of light-toned rocks were reported along Coprates Chasma, Ophir Chasma, Candor Chasma, Oudemans Crater, Trouvelot Crater, etc. (HIRISE). There is striking similarity between the appearance of Martian pebble conglomerates and JS from the epicenter of Tunguska catastrophe (Daily Mail Reporter 2013). The idea of sedimentary origin meteorites was suggested for the first time in late 1930s. We have expressed this idea starting from early 1970s when JS was found; our considerations were published mainly in Russian literature (Anfinogenov and Budaeva 1998). Before us, in 1947, Cross FC reported about three specimens including two grayish fine-grained sandstone found in the United States, that, in his opinion, deserved consideration to be of cosmic origin (Cross 2012). He briefly reviewed account of 1939 from Dr. Assar Hadding, the Director of the Geological Institute in Lund (Sweden), who described two specimens (one of limestone and one of sandstone) that he believed were meteorites (Hadding 1940). In 2000s, international research team did a series of experiments whose objectives were to determine the effects of thermal alteration during atmospheric entry of Martian analogue sediments (Foucher et al. 2009; Brack et al. 2002). In these experiments, the holders with terrestrial sedimentary stone specimens were fixed onto the capsule near the stagnation point of FOTON, robotic spacecraft used by Russia and the European Space Agency for research in the microgravity environment of Earth orbit. Valuable data were eventually generated showing that the admixture of fragments of silicified volcanic sediments and space cement survived the thermal shock well, forming a white fusion crust (Foucher et al. 2009).

**Conclusions.** The following findings confirm high-speed movement of JS through the near-surface permafrost layers: (i) buried slightly arching furrow in the permafrost; (ii) shear-fractured splinters of JS; (iii) SS with its own furrow; (iv) splinters with glassy coatings; (v); half-cleaved pebbles on surfaces of JS and its splinters; (vi) data of reconstruction studies; and (vii) clear consistency in geometry of spacial arrangements of all splinters, furrows, cleaved pebbles, etc. The landing velocity of JS was ≈ 550 *m/c*. Trajectory of JS landing agreed well with trajectory of Tunguska cosmic body flight. Discovery of this sedimentary origin high-silica boulder with signs of high-speed collision with Earth in the epicenter of the Tunguska catastrophe may suggest its cosmic origin and relation to the 1908 Tunguska event. Organization of the international interdisciplinary research group is necessary to identify Martian meteorites of sedimentary origin composed of fine- and coarse-grained silicified conglomerates.

The authors declare no competing financial interests.

Correspondence and requests for materials should be addressed to anfiyj@gmail.

**Supplementary Information**:

Our study provides rationale for (i) lithological and thermoluminescent analysis of glassy surfaces of JS splinters; (ii) isotope analysis of silicon and oxygen in JS; (iii) radioisotope analysis of JS; (iv) comparative analysis of JS, similar terrestrial rocks, and Martian meteorites; (v) paleontological analysis of JS; (vi) exploration survey of the preserved subsurface layers of the ground around JS in the east-south sector; (vii) exploratory excavation of the point where JS penetrated the ground; and (viii) military-criminology analysis of the permafrost destruction caused by JS; the destruction may be similar to that caused by ricochet of a fallen, but unexploded, bomb. All splinters with glassy coverings should be recovered as soon as possible due to imminence of the replacement of the fusion crust by terrestrial minerals.

Table I. Spectrometry data. Elemental analysis of John's Stone.

| **Element** | ***n x* 10$^{-3}$%** |
|---|---|
| Pb | 0.20 |
| Cu | 4.00 |
| Zn | 5.00 |
| Co | 0.10 |
| V | 0.30 |
| Cr | 3.00 |
| Ni | 1.00 |
| Ti | 300.00 |
| Mn | 20.00 |
| Ga | 0.10 |
| Ba | — |
| Mo | 0.20 |
| Be | < 0.10 |
| Sr | — |
| Zr | 15.00 |
| Nb | 1.50 |
| B | 0.60 |
| P | 40.00 |
| Ce | < 0.10 |
| Ag | 0.15 |
| Y | 1.00 |
| Li | 1.00 |
| Sn | 0.10 |

**Figure captions:**

**Fig. 1.** Schematic map of the epicenter area of the Tunguska catastrophe of 1908. Location of John's Stone and penetration funnel-like structures.

**Fig. 2.** Scheme of the exploratory excavations and main findings (1972) around John's Stone. **JS:** John's Stone. **SS:** Satellite Stone which was reconstructed to fit the south part of JS (SS*). **RRH:** "Red-Riding-Hood" specimen found under the south part of SS and reconstructed to fit JS cupola (RRH*). **2—7:** The 0.5- to 2.0-m-deep exploratory trenches. **9:** Large specimen found in 1.7-m-deep trench; this specimen was reconstructed to fit bottom side of JS. **11&13:** Sectors of the furrows, enriched with numerous splinters of JS. **12:** Spot of sand (see also Figure 3 C&D) where two extraordinary specimens with continuous well-formed glassy coatings were found; these specimens fitted each other, but were spaced-apart by 50 *cm* and deposited at different depth. **14:** Fragment of the original furrow generated by movement of the JS&SS precursor. Direction of this furrow apparently resembles trajectory of meteorite landing. Beginning part of the furrow was left undisturbed for further studies.

**Fig. 3.** Photographs (1972) of John's Stone and related findings. **A:** John's Stone and its scaly plate-like splinters (bottom left). **B:** Furrow filled with loose black soil. White arrows show profile of the furrow. **C&D:** Spot of light-toned sand containing extraordinary specimens with continuous well-formed glassy coating. **E:** Satellite Stone which fits the south part of John's Stone.

**Fig. 4.** Example of the reconstruction study showing that Satellite Stone broke off John's Stone precursor. **A:** Photograph of the south wall of John's Stone. Blue tracings show characteristic lines of its superficial architectonics. **B:** Photograph of Satellite Stone. Yellow tracings show characteristic lines of its superficial architectonics. **C:** Blue tracings are taken from the image A. Orange tracings indicate overlay (absolute matching) of characteristic lines from A and B.

**Fig. 5.** Images of five specimens from John's Stone. Photos taken in 2013. Red arrows indicate half-cleaved shear-fractured pebbles. Yellow arrow indicates crack in the crust. **S1F1 & S1F2:** Two views of light-toned specimen 1 with characteristic bowed shape, smoother outer surface, and freshly fractured edges. **S2F1:** Glassy covering on specimen 2. **S3F1, S3F2 & S3F3:** Three views of specimen 3. **S4F1 & S4F2:** Two images of flat plate-like splinter (specimen dimensions: thickness of 1 to 4 *cm*; longitudinal size of 20 *cm*). **S5F1:** General view of flat plate-like splinter 5. **S5F2:** Signs of surface weathering and rust flowering on the crust of splinter 5. **S5F3:** Reverse side of splinter 5. **S5F4.** Magnified image of splinter 5.

Fig. 1

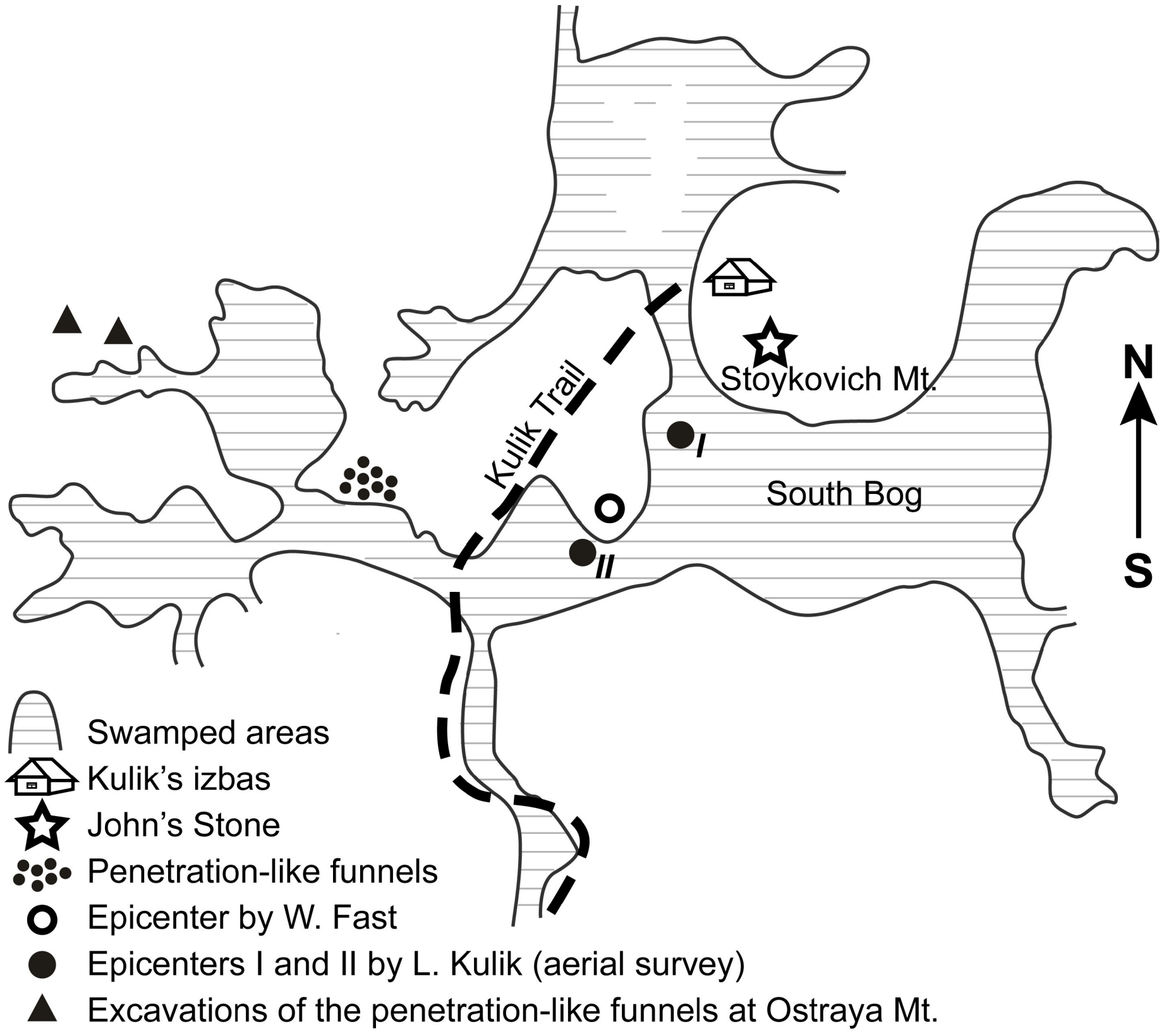
Kulik Trail
Stoykovich Mt.
South Bog
I
II
N
S
Swamped areas
Kulik's izbas
John's Stone
Penetration-like funnels
Epicenter by W. Fast
Epicenters I and II by L. Kulik (aerial survey)
Excavations of the penetration-like funnels at Ostraya Mt.

Fig. 2

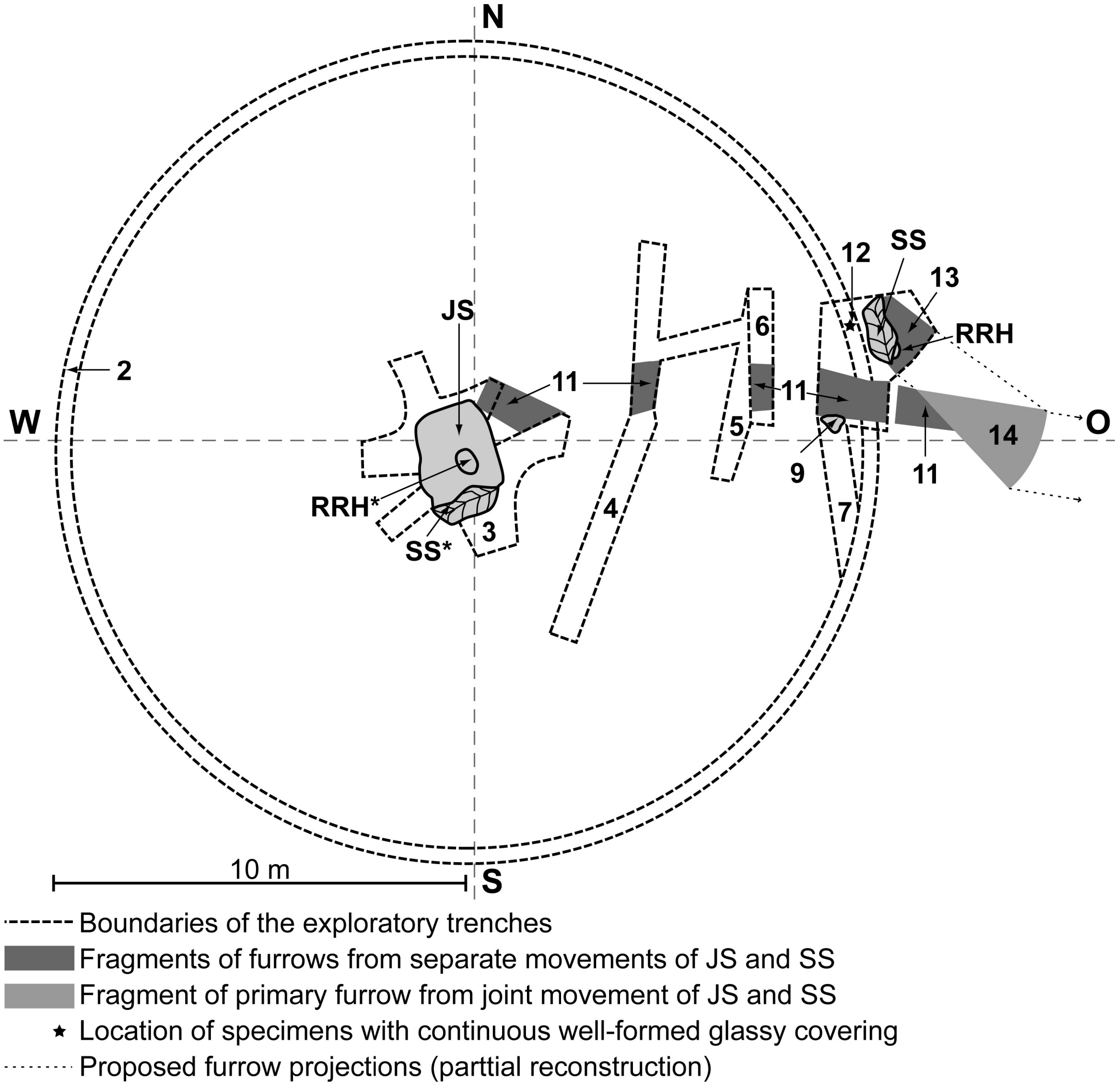
N
W
O
S
JS
SS
RRH
RRH*
SS*
2
3
4
5
6
7
9
11
12
13
14
10 m
Boundaries of the exploratory trenches
Fragments of furrows from separate movements of JS and SS
Fragment of primary furrow from joint movement of JS and SS
Location of specimens with continuous well-formed glassy covering
Proposed furrow projections (parttial reconstruction)

Fig. 3

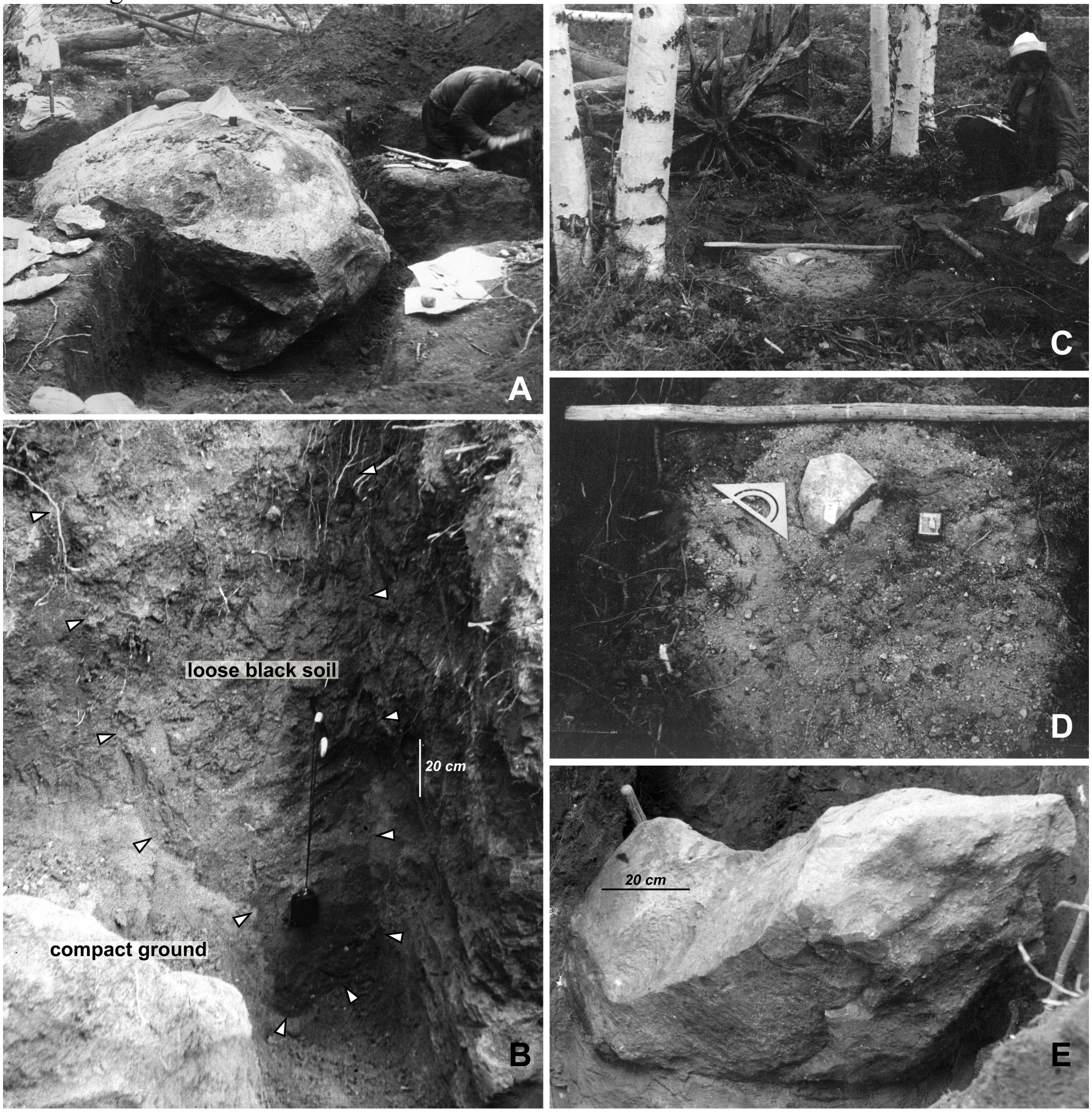

Fig. 4

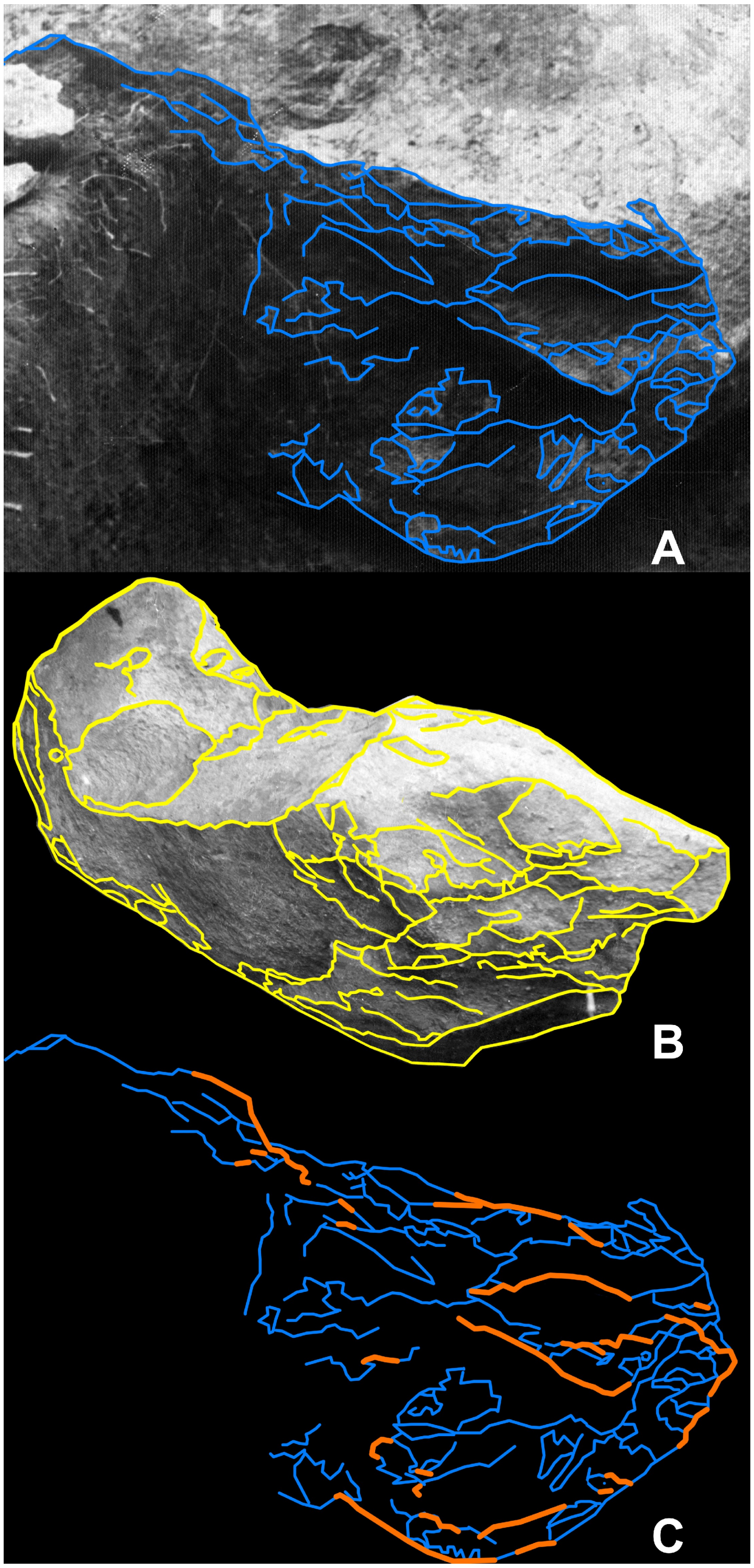

Fig. 5

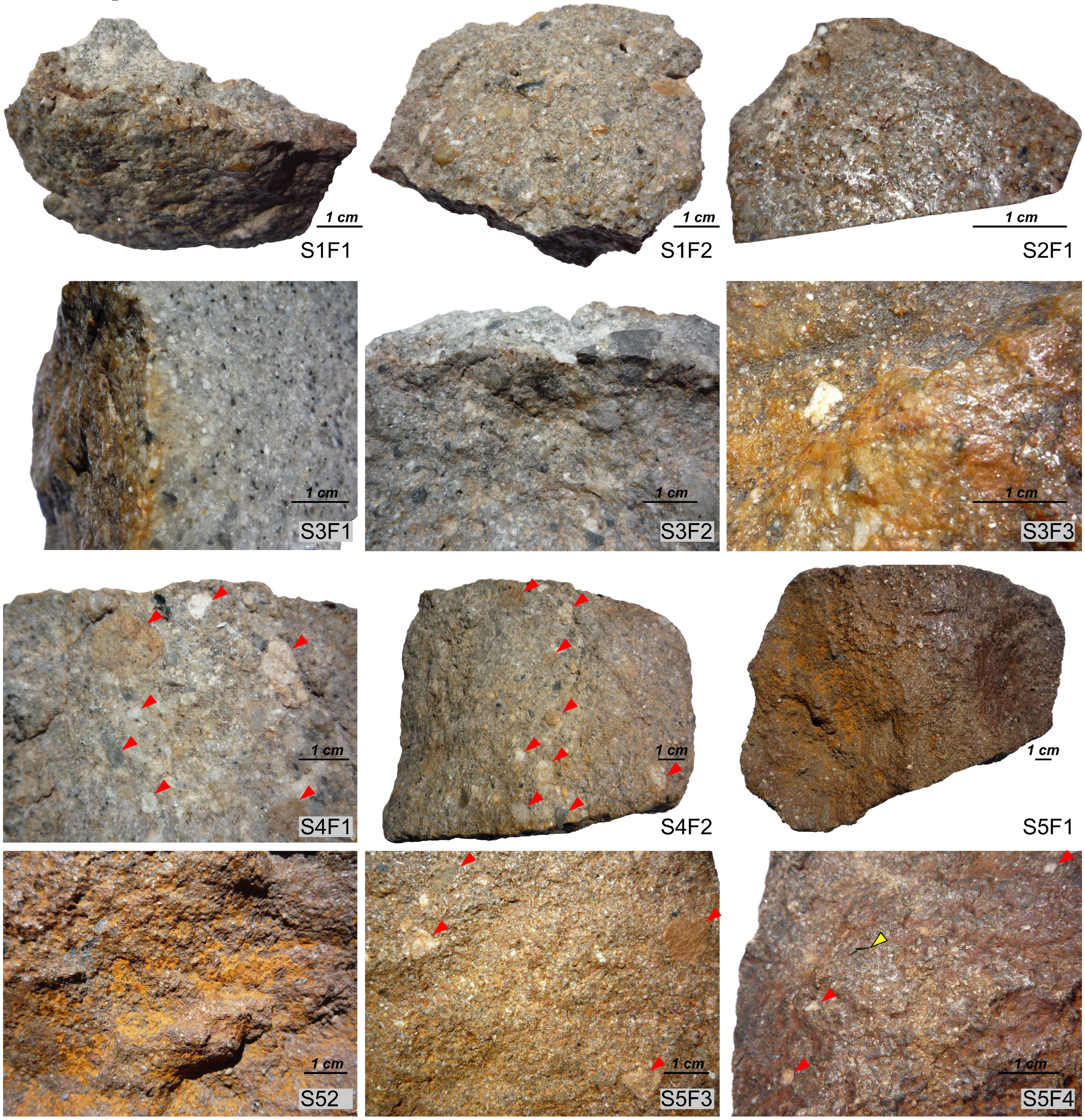